# Identifying interception possibilities for WhatsApp communication

Dennis Wijnberg[1], Nhien-An Le-Khac[2*]

*[1] National Police of the Netherlands, The Netherlands*
*[2]School of Computer Science, University College Dublin, Ireland*

**Abstract**

On a daily basis, law enforcement officers struggle with suspects using mobile communication applications for criminal activities. These mobile applications replaced SMS-messaging and evolved the last few years from plain-text data transmission and storage to an encrypted version. Regardless of the benefits for all law abiding citizens, this is considered to be the downside for criminal investigations. Normal smartphone, computer or network investigations do no longer provide the contents of the communication in real-time when suspects are using apps like WhatsApp, Signal or Telegram. Among them, WhatsApp is one of the most common smartphone applications for communication, both criminal as well as legal activities. At the beginning, WhatsApp communication between smartphone and server used to be in plain-text and therefore visible to law enforcement during a wiretap. Early 2016 WhatsApp introduced end-to-end encryption for all users, immediately keeping law enforcement officers around the world in the dark. Existing research to recuperate the position of law enforcement is limited to a single field of investigation and often limited to post mortem research on smartphone or computer while wiretapping is limited to metadata information. Therefore, it provides only historical data or metadata while law enforcement officers want a continuous stream of live and substantive information. This paper identified that gap in available scenarios for law enforcement investigations and identified a gap in methods available for forensic acquiring and processing these scenarios. In this paper, we propose a forensic approach to create real-time insight in the WhatsApp communication. Our approach is based on the wiretapping, decrypting WhatsApp databases, open source intelligence and WhatsApp Web communication analysis. We also evaluate our method with different scenarios in WhatsApp forensics to prove its feasibility and efficiency. Through these scenarios, we found that by providing real-time intelligence such as profile pictures, their activity, voice and video call behaviour including location data as well as remote access to a suspect WhatsApp account, their conversations including voice messages, (live) geolocation, shared contacts, documents, images and videos are made accessible. Hence, our corresponding method can be used by law enforcement agencies around the world to reinforce their position in the world of WhatsApp communication interception.

## 1. Introduction

In the early days of interception there was not much of a challenge for law enforcement. Wiretapping was easy and sufficient since no communication line was encrypted, except for military lines. The same goes for Short Message Service (SMS) traffic that is not encrypted and therefore easy to reconstruct for law enforcement, who had a real-time intelligence position on the suspect and the investigation. But, SMS was a pay per message service with 140 bytes of size for each message.

When alternative messaging applications were introduced like WhatsApp in 2009 and the adoption increased, the amount of messages grew substantial. Messaging was now free and virtually unlimited in size. SMS was degraded in terms of usage, but there still was not much of a problem, since WhatsApp was in plain-text XMPP [1]. Later, when WhatsApp implemented their end-to-end encryption in early 2016, there was an unbreakable barrier introduced in the readable interception of WhatsApp messages. With 1.5 billion active users and 60 billion messages each day, WhatsApp created a giant gap in the law enforcement intelligence position as the downside of secure communication for the public.

On a daily basis, law enforcement investigators still find themselves unable to intercept encrypted communication, to a readable level, between suspects in organised crime groups. This includes WhatsApp communication since this is the most popular messaging app nowadays. Since the seizure of Ennetcom [2] and PGPSafe [3], the architecture for private messaging is considered to be a risk for criminals and regular apps like WhatsApp, also providing end-to-end communication. Especially now several suspects have been sentenced to life [4] based on this information.

---

[*] Corresponding author
Email addresses: Dennis.Wijnberg@klpd.politie.nl (Dennis Wijnberg)
an.lekhac@ucd.ie (Nhien-An Le-Khac)

In January 2015, when Web WhatsApp was introduced, this opened a new opportunity to use WhatsApp for users and a new platform to investigate. Since communication was still unencrypted until early 2016, law enforcement could simply wiretap the suspect. However, since WhatsApp has implemented their end-to-end encryption most researches were done on artefacts and methodology Web WhatsApp [5, 6, 7]. These researches were all limited to post-mortem investigations. However it provided insight in which forensic artefacts could be found on a computer. Additionally research was conducted on the discovery of artefacts in wiretapped data during WhatsApp phone calls [8], because data streams containing these phone calls are also encrypted.

However, there still is a gap in knowledge, since post-mortem analysis on computers and smartphones [9] can only provide historical data and wiretapping [10] also does not provide real-time content of the conversation. In addition, existing methods for forensic investigation do not meet the needs and requirements in the investigation because they do not anticipate on a continuing data stream.

Hence, this paper aims to fill that gap by investigating which alternative methods for interception could be identified by combining existing knowledge and new experiments on existing WhatsApp functionality. By research on the security features that are optional and disabled by default, insight was acquired in the exploitation of these features. We also evaluate this research through scenarios. For each scenario the information, acquiring process and risk of discovery by the suspect are described and how law enforcement operatives can minimise the chance of being vulnerable from this attack.

Five (of which four disclosed) working scenarios were found in which a combination of techniques were used. During two scenarios of which one is disclosed, it was possible to have remote and complete insight in all WhatsApp chats, including all media and (live) location. During other methods it was possible to either retrieve metadata information (picture, online status, name, IP address) or to take over the current WhatsApp and gather the names of group chats, the suspect of the group and its members and messages that were either in transit (both one-on-one as well as group-messages) or after the takeover. This allows law enforcement officers also to impersonalize the suspect and send messages on their behalf.

Existing methods for forensic investigation do not meet the needs and requirements in the scenarios identified because they do not anticipate on a continuing data stream, existing methods were expanded accordingly.

Each scenario concludes with how the scenario can be prevented by the suspect, this includes law enforcement operatives to become a victim of counter surveillance program either by foreign state actors or organised crime gangs.

The rest of this document is structured as following: Section 2 reviews related work in literature. We describe our methodology in Section 3. We present our experiments in Section 4. Finally, Section 5 shows the conclusions of this research and open questions for future work.

**2. Related work**

Most forensic work regarding WhatsApp is done on mobile phones. Forensic software such as Cellebrite UFED often has the possibility decrypt databases on the phones in to readable reports. However, it is not self-evident that this succeeds. Encryption on mobile phones gets better or other forensic countermeasures are taken. When placing the information centric instead of the device, smarter methods can be used to get the information the investigator needs. This includes the use of Web WhatsApp to retrieve WhatsApp data.

Yudha et al. (2017) identified the possibility to find browser artefacts and their location in post-mortem investigations [5]. The model helps researchers to investigate Web WhatsApp in several browsers (Chrome, Firefox, Safari and Internet Explorer), but included no room for live-data forensics.

In 2019 Vukadinović [7] did research on which artefacts, if any, of Web WhatsApp (and the native client) could be further recovered and identified. It showed that all messages send (40), nearly all deleted messages (2/3) and all pictures send (3/3) could be recovered in both Chrome as well as Firefox. It is in line with expectations that when a larger dataset is used, results cannot remain this good because of lazy loading [11] and the limited amount of storage available through the browser [12, 13]. But in his research, the author focused on post-mortem investigations on computers, the focus was never on intercepting real time information.

Buchenscheit et al. (2014) discuss the privacy implications of the 'precense' feature data and how it can be used to analyse activities and routines [14]. Kloeze (2017) showed that acquiring this data can be automated by using the Javascript API from Web WhatsApp [15] and an enumeration to scrape information from large amounts of WhatsApp users.

Real time information can also be acquired by looking at WhatsApp as a social media platform instead on limited to instant-messaging. Since the acquisition of Instagram by Facebook in April 2012[a] and WhatsApp by Facebook in February 2014[b], similarities and overlap cannot be denied. Instagram implemented direct messaging, Facebook as well as WhatsApp have implemented "Statuses" (which was originally an Instagram "Stories" feature) and Facebook is currently implementing end-to-end encryption in their Messenger service (not by default)[c].

Since WhatsApp cannot function without a network (cellular or WiFi) connection, WhatsApp network forensics can provide real time data as well. Before WhatsApp embedded end-to-end encryption, the interception of these messages was less of a challenge [1].

Sgaras et al. [16] focused on the forensic acquisition and analysis of different instant messaging applications such as WhatsApp, Viber and Tango on both iOS and Android platforms. Discussions were also made on IM chat cloning and communication interception. Again, the authors focused on post-mortem investigations.

Quick et al. (2016) describe that data from regular smartphone and computer extractions can be combined with open source intelligence such as discovered by Buchenscheit to add value to the data, enabling to greater understanding of the criminal environment [17].

In [18], authors proposed a volatile memory forensic approach for the retrieval of social media and instant-messaging evidences for some apps such as Skype, iMessage, Facebook Messenger Viber and WhatsApp. However, this approach is not for a continuing data stream.

Not only the scenario in which case live data (regardless of the form) can be acquired, methods of investigation is also of interest. Actoriano et al. (2018) proposed a forensic framework for Whatsapp Web while investigating WhatsApp [6]. This method was intended for post-mortem investigations.

## 3. Methodology

While law enforcement wants real-time intelligence from suspects (meta-data and the conversations), digital forensic examiners focus on the investigation of seized devices.

As described in Section 2 above and visualised in Table 1, there is a gap in research in on how law enforcement can retrieve future or real-time intelligence of WhatsApp communication and which model must be applied by the forensic digital investigator.

A regular wiretap shows only encrypted packages and within these packages are end-to-end encrypted messages on a per message basis using a chat key. WhatsApp does not store the decrypted messages nor the private keys to decrypt them. Therefore requesting these keys or messages will not contribute to the primary objective. Although WhatsApp does cooperate with law enforcement, but because of their "privacy by design" on the server end, it simply cannot cooperate in decrypting those messages. In addition, WhatsApp cannot be questioned for information (too specific), since the unintended use of their software might cause patches that decrease the possibilities that are found in our research.

| Method | Meta Data | OSINT Data | Historical Content | Future Content | Real-time Content |
|---|---|---|---|---|---|
| **Tsai [8]** | √ | | | | |
| **Vukadinovic [7]** | | | √ | | |
| **Buchenscheit [14]** | | √ | | | |
| **Kloeze [15]** | | √ | | | |
| **Actoriano [6]** | | | √ | | |
| **Yudha [5]** | √ | | √ | | |

Table 1. Comparison of current approaches in literature

As our research objective is how existing WhatsApp functionality can be used for law enforcement investigations, we propose a method for conducting the investigation. The most significant difference with existing methods is the existence of an iterative sub-process regarding the continued acquiring, analysing and documenting of the data. The common stages prepare, acquire, analyse and report are adopted and expanded (Figure 1).

*3.1 Prepare*

This stage includes three main tasks as follows:
- (1.1) Notification is the start of the investigation. The trigger to start. This can be a filed report or an event such as a murder or an intercepted drugs shipment. This includes creating the research questions such as: "What information do I need for my investigation?" and "What options do I have to get the needed information?"
- (1.2) Requesting the needed legal information in order to be allowed to conduct the intended investigation.
- (1.3) Preparation such as determining the make and model of the items that is subject to the investigation, hardware and software that is needed and people that are needed to perform the investigation.

*3.2 Acquire*

In some scenarios, the smartphone (or any device) is not physically needed. In that case, information can be acquired remotely. The following steps can be conducted:
- (2.1.1) Creating a clean sterile environment avoids data contamination. For example a clean virtual machine. If the system is connected to the internet, the 'digital crime scene' must also be secured from remote intruders. An up-to-date anti-virus or firewall (including intrusion detection and prevention) contributes to that goal.
- (2.1.2) Before the data is acquired, a solid baseline must be stored this includes information about the start of the investigation, methods used, the officer conducting the investigation and steps taken to reduce interference and contamination.
- (2.1.3) The method of choice is conducted. This includes the methods such as mentioned in chapter five.
- (2.1.4) The method of choice can either produce nothing (unsuccessful), a one-time dataset or a continuing flow of real-time data.
- (2.1.5) In case of a continuing flow of real-time data, the data from this flow must be preserved. This includes shielding this information to a "sneakernet".
- (2.1.6) As the data keeps flowing, the data must be captured and logged as real-time as possible in order to prevent interference.
- (2.1.7) The storage process is, especially in case of real-time data, intensive. The device acquiring the data must be fault-tolerant as well in terms of power as in the event of disk failure and the device must have sufficient storage capacity and must be fast enough to store the data. The data must also be copied to a second location. This can either be off-site (recommended in terms of data safety, but the data transfer is also a risk).

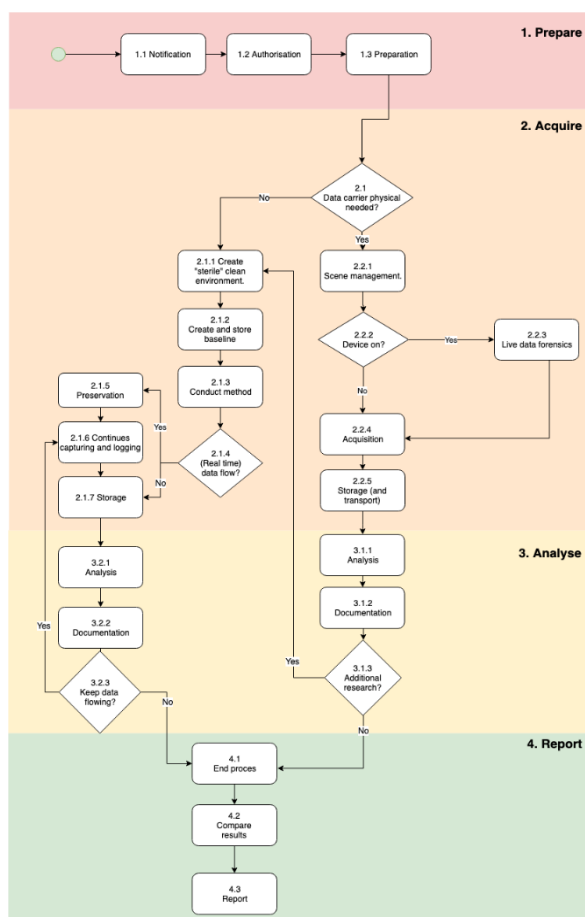

Figure 1. Proposed forensic process

When the device is physically needed, we hence follow these following steps:
- (2.2.1) Scene management; this includes securing the scene and documenting the scene.

Both the physical and the digital crime scene. This also includes handling the suspect holding the phone. Separating the suspect from his phone, can affect the chances of accessing the phone, i.e. a smart watch, biometric access, etc.
- (2.2.2) If the device is still on: (2.2.3) Live data forensics can be applied, this includes: avoiding drainage of the battery, accessing and acquiring the volatile data, communication shielding.
- (2.2.5) Acquisition of the device. This can be done on the scene for tactical reasons (wiretap, friendly treatment in order to benefit during a hearing) or the device can be seized and brought back to the lab.
- (2.2.6) For the storage of the acquired data (cf. point 2.1.7 above). Transport should always be done according to the rules of chain of custody and chain of evidence which apply to the jurisdiction. If the device is on, sufficient power must be guaranteed.

*3.3 Analyse*

- (3.1.1 / 3.2.1) The data is analysed in this stage. This can be done through several methods such as OCR, after which automated (machine learning to detect identity documents, weapons, suspects) analysis or manual analysis can be done.
- (3.1.2) Documenting not only the outcome of the analysis but also the input and the process. If no additional research is done, this will be the source for the report as mentioned in Section 3.4.
- (3.1.3) Is additional research needed to get more relevant data that contributes to the goal of the investigation?
- (3.2.2) Documenting not only the outcome of the analysis but also the input and the process. These are preliminary findings and can be used as attachments or input for the report (cf. Section 3.4).
- (3.2.3) If the process should be continued because the information that was searched for is not yet discovered or exculpatory evidence is found, the process proceeds at step (2.1.6).

*3.4 Report*

- (4.1) Ending the process requires, besides juridical and tactical approval, a technical thought. Depending on the scenario and methods used, it is possible to resume the interception if the process was ended in a proper way. Also, the information needed to resume should be classified as strictly confidential.
- (4.2) Comparing results between different iterations of requesting the information can both proof the consistency of the results as well as show differences in those versions. Differences must be explained. It makes it harder to dispute the validity of the evidence in court.
- (4.3) Report about the several stages. The preliminary findings from the previous stage are bundles into a consistent and logical report that is complete and clear on the research questions.

## 4. Experiments

In order to evaluate our approach, four experiments were carried out based on scenarios.

*4.1 Experimental platform*

Data used was divided in two groups: audio and text. In case of a conversation a random song from Spotify was used to generate traffic. In the case of text Test data (not random) was generated from https://www.lipsum.com in English. The creating accounts in the operating system of the phone as well as the data used for creating the WhatsApp accounts are showed in Table 2.

|  | Apple iPhone 8 | Samsung Galaxy S8 (SM-G950F) |
|---|---|---|
| Name | Mark Boerboom | Mark Boerboom |
| E-mail | markboerboom@icloud.com | markboerboom73@gmail.com |
| Number | 0031621444833 | 0031621440487 |
| OS | iOS 12.1 | Android version 7, 8, 9 |
| WhatsApp version | 2.19.80 | 2.19.216 |

Table 2. Experimental platform

For forensic acquiring both phones Cellebrite UFED4PC was used. For the investigation UFED Cellebrite was used combined with SQLiteBrowser and sqlite as a command line tool. For forensic acquiring and investigation process on the computer (virtual machine), Forensic Toolkit (FTK) was used to image and extract the files needed. This was later repeated with Kali Linux (live boot), where the partition containing the browser data was mounted as read-only. This was done to prove the ability for selective imaging. In order to live boot, secure boot was disabled. Further, WhatChanged Portable was used to track changes within the operating system as a part of live data forensics and establishing what changed inside the operating system.

In all software including the operating systems of the phones, the option of automatic updates was disabled. The internet connection was available over WiFi (to a Huawei 4G LTE hotspot), except for one scenario in which the Macbook needed a wired connection in order to use the WiFi-radio as an access point.

*4.2 Experiment 1- Take-over the WhatsApp account*

The first scenario is to take-over the WhatsApp account in the legal framework. In this scenario the (re)activation message was intercepted (wiretapping, reading along in a message preview or if the SIM is seized) via the following steps:

1. SMS traffic should be intercepted (wire-tapped) at the telecom provider or that SMS messages can be read (nearly) real-time.
2. WhatsApp needs to be installed on the suspect's phone.
3. When the number was entered as if it was the researchers own number, WhatsApp sends a SMS message containing a verification code.
4. After the verification code was entered as seen in the intercepted SMS into the suspect's phone, the WhatsApp account was taken over on the new device.
5. Due to this activation, the phone on which WhatsApp was previously registered with this number, is now disconnected to the account.

In this way the WhatsApp account was taken over. The following information can be acquired in this scenario: Information about WhatsApp groups the suspect was subscribed to (including group names), "in transit" WhatsApp group messages that were not yet received on the suspect's phone, phone numbers of members of the group, not including their display names as long as they have not send a message, but including their display names if there was a message "in transit" to the group and messages that were "in transit" 1 on 1, including display names of their sender.

The risk of discovery is inevitable. The suspect will be logged off by WhatsApp on the original phone and therefore be alarmed that something is off. In the meantime the messages can be received. Hence, the researcher need to choose a time frame in which WhatsApp is not used by the suspect; during a flight, night time or from another time frame learned from WhatsApp Open Source INTelligence (OSINT).

Besides, if the third party (the sender of messages to the suspect) had "security notifications" enabled, this will show a warning on their end as shown in Figure 2.

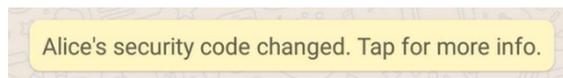

Figure 2. A security notification when WhatsApp is activated on another phone

The use of this method can be prevented by using "two step verification" the WhatsApp account [19]. This way WhatsApp enabled the suspect to enter a six digit PIN and an email address as well. If the PIN cannot be entered and there is no access to the suspect's email, this scenario cannot be completed.

*4.3 Experiment 2- Creating a WhatsApp web session on the suspect's phone*

Since WhatsApp introduced their Web service in January 2015, they allowed users to use their service from another device than their phones. A session can be created when the investigator has access to WhatsApp on the suspect's phone. In this scenario, the phone, therefore needs an active internet connection to be able to communicate with the web application. The WhatsApp web service functions as an intermediary and communicates over web sockets (Figure 3).

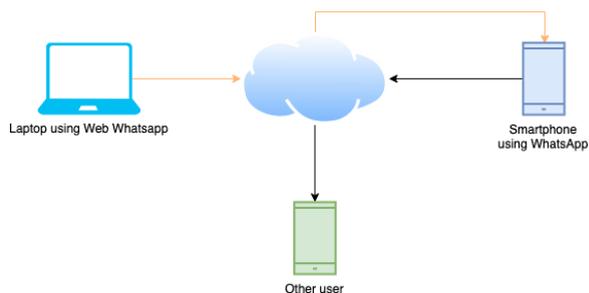

Figure 3. Schematic overview of connections using Web WhatsApp

A periodic QR-code can be scanned and a pair is created between the devices for as long as the session was not explicitly removed or limited in use. As soon as the pair was made, the computer using Web WhatsApp downloads the list of conversations, their members and images and the list of blocked users that need to be loaded to proper serve the viewport of the current screen.

Using this scenario, all WhatsApp data that is on the phone can be retrieved. This included contacts, their profile pictures and media (photos, videos, audio fragments and other documents) that were either sent (successful and unsuccessful) or received on the suspect's phone. However, phone calls are not visible from Web WhatsApp and neither is its content (recording) accessible, since it was not recorded. Another exception is the messages that were explicitly removed from the device and where the notification that the message was deleted, was removed either.

Besides, depending on the make and model of the phone, the suspect can get a notification during or shortly after the use of a web session [20]. Also, if the content of the WhatsApp on the phone is downloaded at once, this can drain the battery on the phone. The last risk identified is within that messages that are read, cannot be marked as unread. This can be done on a chat basis, but not on a message basis.

This method can be prevented by adding extra biometric security to the WhatsApp app on the suspect's phone. This creates a new barrier of security before creating a session [21].

*4.4 Experiment 3- WhatsApp phone call*

A WhatsApp phone call can also provide new insights. WhatsApp chose to encrypt phone calls as well [8]. Therefore the content of the conversation cannot be retrieved. However, the IP address of both parties can be retrieved. This can provide insight in the network of the suspect and can show activity in which person calls who and when. The IP addresses can be looked up and trace back to an internet connection elsewhere.

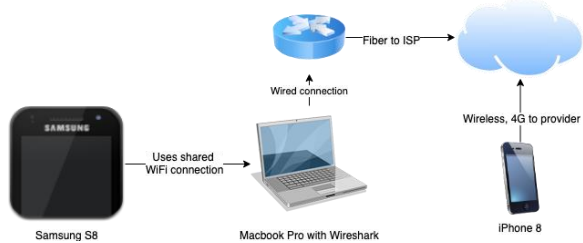

Figure 4. Schematic overview of the interception

The protocol was designed to use the WhatsApp servers as a routing service in order to send the push notification and establish the connection between the caller and the called. As soon as the connection is established, the WhatsApp server intermediates in establishing a connection between the two phones as shown in Figure 4.

The IP addresses used in the lab environment shows the Samsung phone (local IP 192.128.2.2) calling the iPhone using a 4G cellular connection (external IP 62.140.137.15). The size of the package is 86 Bytes and the protocol used is STUN. This is an outgoing WhatsApp call flow through intercepted network traffic (Figure 5).

The setup used to investigate the traffic was designed to intercept all the traffic from the inspected device (i.e. the Samsung). The same method was repeated for both devices and for both incoming and outgoing calls.

| No. | Time | Source | Destination | Protocol | Length | Info |
|---|---|---|---|---|---|---|
| 155 | 23.990789 | 192.168.2.2 | 62.140.137.15 | STUN | 86 | Binding Request |
| 166 | 24.596166 | 192.168.2.2 | 62.140.137.15 | STUN | 86 | Binding Request |
| 184 | 25.205671 | 192.168.2.2 | 62.140.137.15 | STUN | 86 | Binding Request |
| 197 | 25.808196 | 192.168.2.2 | 62.140.137.15 | STUN | 86 | Binding Request |

Figure 5. Intercepted network traffic

Using a display filter `stun.type == 0x0001 and ip.dst!=192.168.1.0/16`[d], the traffic can be differentiated from the rest and internal packages are filtered as well. Assumed is that the internal network uses 192.168.X.X private IP space. Still, some other packages can be displayed that do also match this pattern. If multiple calls were made to different parties during the capture period, all will be displayed. The same behaviour is seen is cases where a handover was done.

This method will give the investigator insight whether the suspect is using WhatsApp for phone calls and if so, the IP address of the person the suspect is calling with. If the person on the other end is on a home WiFi-connection, the IP address of their location is then known, unless that other party is using forensic countermeasures such as a VPN-connection.

In the person on the other end is using a cellular connection, carrier grade NAT is used. Therefore the subscriber can only be identified if the network provider has a log file port numbers, IP addresses and time. In the Netherlands, for example, this is stored for a few days, depending on the provider.

If the subscriber is anonymous, for example in the case of non-registered pre-paid, topping up transactions (wire transfer, credit card) might give the insight needed to identify the subscriber after all.

---

[d] This should be changed into the private network of choice.

It must be noted however that this method will most likely can only produce a limitative view of the WhatsApp calls since all connections the subject is using, must be lawfully wiretapped. When the subject is using public Wi-Fi, this would not be opportune, unless the amount of users is limited, the suspect is a regular user of the connection and other methods or connections remained unsuccessful.

Further, if two suspects are wiretapped at the same time, conversations can be proven by the fact that the beginning and the end of conversations (WhatsApp phone calls, but messages nonetheless) can be recognized in the data streams on both ends.

In this method, the data was captured on a network gateway that contains only traffic from the Samsung S8 to the Macbook Pro and the other way around. Another method could be to capture all data and limit the data using Wireshark's display filters. Data is captured in a raw format and stored as .pcap files. These files can be analysed using WireShark or commercial systems for law enforcement.

Tools and platforms like Shodan.io that gather open source information from port scans, trace routes and network registries such as RIPE can help to make a selection which IP addresses are interesting for further analysis.

Depending of the country of origin of the IP, the IP address can be looked up in a registry or at the internet service provider to track down the corresponding subscriber. In case of a public Wi-Fi it will be harder to track this to a certain person.

This method can only succeed if both parties are in each other's contacts list. Otherwise STUN traffic will not provide the IP address of the other party.

This method can be prevented by calling numbers with WhatsApp that are not stored in each other phones. This might seem impossible since the user interface does not supply a screen to enter a number. But a suspect is able to call a WhatsApp group member that is not in the contact list.

*4.4 Experiment 4- WhatsApp OSINT*

WhatsApp can also be considered a social media platform and the privacy settings are not extremely tight by default. Last seen, profile photo and about-message can be seen by "Everyone". The last experiment described in this section can be considered to be an OSINT method that can be used to identify a phone number.

The concept is as simple as effective: add the phone number to the contact list of the investigators phone and, after a couple of seconds, open a WhatsApp conversation. This will show the profile picture of the subject. Using reverse image tools, such as Google Reverse Image search, Yandex or TinEye, these images can lead to the subject. If a set of pictures of person of interest in a case exist, faces can be compared using Microsoft Azure Face API in order to score the given pictures on similarity.

Depending on the settings as configured by the suspect: the profile photo, the "*last seen date and time*" and the "*about*" the subject can be visible to anyone.

If the risk can be taken to send the subject a message, their display name becomes visible as soon as they answer the message. Depending on the subject an advertisement on the internet or a Tinder or Grindr match can be an excuse for a "*wrong number*" excuse.

In order to get the profile image in the highest quality, Web WhatsApp can be used on the investigators account in order to download the profile picture and report about this.

In example, by enumeration large amounts of phone numbers and profile pictures can be gathered using the WhatsApp JavaScript API [22] and earlier [14] discussed that it can be used to detect communication partners.

The information that can be acquired is limited to a last-seen date (and therefore activity) and a profile picture. The profile picture can be source for further investigation such as reverse image search. If the investigator can send a message to the suspect on which the suspect replies, the display name of the suspect can also be acquired.

*4.5 Discussion*

The purpose of this paper was to explore the possibilities for law enforcement in regard to WhatsApp. During this research the possibilities of alternative methods for law enforcement with WhatsApp are explored and several are found. In the most comprehensive situation a scenario was found that would enable law enforcement to read and hear along everything and would work regardless of the circumstances and without the suspect being able to detect.

The scenarios and methods proposed provide a way for forensic investigators to acquire live WhatsApp data during an investigation. Some scenarios are based on unintended behaviour of WhatsApp.

The government has a virtual social contract with its citizens. The government has to protect its citizens from cyber-attacks but it has also the need to execute attacks their selves during a criminal investigation or intelligence process. This is an ethical discussion where, depending on the weight of each of the items, both can be argued.

Also, the trustworthiness of information can be discussed. For example, the Minister of Justice of the Netherlands stated in an interview [23] that he wants to talk to companies such as WhatsApp/Facebook that provide end-to-end encrypted communication to create a decryption key for the messages in their software. However seems understandable from a prosecution point of view, it is not in the best interest of the public. In other countries the bar for government interference is not as high as in the western society where judges have to approve such a breach of someone's privacy. If there is a backdoor, there is a backdoor for everybody and it can also be used by less trustworthy (state) actors.

Last but not least, law enforcement officers can also be attacked by these scenarios. For each scenario a counter-option was also tested and mentioned.

## 5. Conclusion and Future Work

Adding end-to-end encryption to WhatsApp was a huge benefit for all their users, regardless of their intensions. It added privacy to the most popular messaging application. Unfortunately for law enforcement, this also came with the downside of no longer being able to intercept the contents to a usable level.

This research shows that there are still several methods left for law enforcement and intelligence, most of which due to "usability first" instead of "privacy first" paradigm in WhatsApp user experience that created a void for law enforcement. Law enforcement officers were no longer able to read along the content of WhatsApp messages. However, the scenarios and method presented in this research partially recuperate the position of law enforcement agencies. But this problem was never on its own, the growing amount of encrypted data traffic forces researches to think of new scenarios and methods of intervention in criminal activities. A simple wiretap does no longer provide the needed insight.

The research done on WhatsApp can also be done on apps like Telegram, Signal, Threema, Viber, Wickr or BlackBerry Messenger or less common providers such as EncroChat or SkyECC.